\documentclass[pre,preprint]{revtex4}

\usepackage{amsmath,amssymb,amscd} 
\usepackage{latexsym}  
\usepackage{dsfont} 
\usepackage{color} 
\usepackage{rotating} 
\usepackage{epic} 
\usepackage{xspace}  

\newcommand{\be}{\begin{eqnarray*}} 
\newcommand{\ee}{\end{eqnarray*}}   
\newcommand{\ben}{\begin{eqnarray}} 
\newcommand{\een}{\end{eqnarray}}   
\newcommand{\pstex}[4]{
\begin{figure}[#1]
\begin{center}
\includegraphics{#2.eps} \caption{#3} \label{#4}
\end{center}
\end{figure} }
\newcommand{\bra}[1]{\langle  #1 |} 

\newcommand{\ket}[1]{| #1 \rangle}

\newcommand{\dm}[2]{\ket{#1}\bra{#2}} 
\newcommand{\Ref}[1]{(\ref{#1})} 
  
\begin{document} 

\title{Heralded single-photon generation using imperfect single-photon
  sources and a two-photon-absorbing medium}

\author{Thomas Konrad\footnote{the corresponding author:
thomas.konrad@uni-konstanz.de}}
\author{Artur Scherer}
\author{Michael Nock}
\author{J\"urgen Audretsch}
\affiliation{Fachbereich Physik, Universit\"at Konstanz, D-78457
  Konstanz, Germany}
\author{}

\begin{abstract}
We propose a setup for a heralded, i.e.\ announced 
generation of a pure single-photon state given two
imperfect sources whose outputs are represented by
mixtures of the single-photon Fock state  $\ket{1}$ with the vacuum
$\ket{0}$. Our purification scheme uses beam splitters, photodetection
and a two-photon-absorbing medium. The admixture of the vacuum
is fully eliminated. We discuss two potential realizations of the scheme. \\


\end{abstract}
\maketitle

\section{Introduction}

Light is an optimal candidate to be a carrier of quantum information
because photonic states are durable due to the generally weak
interaction with the environment. For this reason,  the only realistic proposals
for long distance quantum communication are  still based on optical systems
\cite{HNDP02,GCTC05,MRTZG03}.   
Photonic states are conveniently manipulated by means of
linear optics and photodetection. However, most schemes
to  process photonic quantum information by means of linear optics
require the ability to produce single photons.
 For example, the ingenious linear optics quantum computation 
scheme proposed by Knill, Laflamme and Milburn 
in~\cite{KnillLaflammeMilburn2001} as well as the  schemes for
  secure quantum 
communication in \cite{Bennett1984,Bennett1992}  rely on the 
availability of single photons.  Single photons entangled with
  the vacuum ($(\ket{1}\ket{0}+\ket{0}\ket{1})/\sqrt{2}$) by a
  symmetric beam splitter  can also be
  employed to teleport and transform single-rail optical qubits
  \cite{Babichev2001,Berry05}.
Moreover, single-photon sources   
are a necessary resource in photonic quantum state
engineering~\cite{ScheelKnight2003,Welsch1999,Babichev2004}. 

In recent years a variety of implementations for single photon sources
has been investigated. Among them are schemes based on single molecule or atom
excitation \cite{MoernerNJP04,AlleaumeNJP04,RempeNJP04}, single ions
trapped in cavities \cite{WalterNJP04}, color centers in diamonds
\cite{Weinfurter00,Grangier02}, quantum dots \cite{Yuan02,
  SantoriNJP04}  and parametric down conversion (PDC)
\cite{Fasel04,Pittman04}.
These sources differ in the wavelength and purity of the emitted
photons, their repetition rate and whether they produce a photon on
demand or heralded, i.e, announced by an event. The latter is for
example the case with PDC-sources. PDC produces randomly  
photon pairs and the presence of one  photon is
indicated by the detection of the other.   

None of the existing single-photon  sources, 
however, emits a pure single photon 
at a given time with certainty. The emission of multiple photons is negligible for
most single-photon sources, cf.\ for example~\cite{WalterNJP04}. Therefore their output in a certain mode
can be modeled by a mixture of a single photon Fock state $\ket{1}$ and vacuum
$\ket{0}$: 
\begin{equation}\label{imperfect}
  \rho = p\ket{1}\bra{1}+(1-p)\ket{0}\bra{0}\,,
\end{equation}
where $p$ is called the \emph{efficiency} of the single-photon source. Good
sources have efficiencies of $p\approx 0.6$. To the best of our
knowledge the highest efficiencies reached so far are $p=0.83$
\cite{Pittman04} and $p=0.86$ \cite{MoernerNJP04}. 

In this article we propose a scheme to process two outputs from
imperfect single photon sources (with efficiency $p<1$) in order to obtain a single photon with
certainty ($p=1$) as desired for quantum information
applications. These outputs may in fact originate from the same
source. Using only linear optics and photodetection, it is impossible 
to produce a pure single photon with certainty by processing 
the outputs of imperfect sources. This is the case 
even if instead of two a finite number $n$
of imperfect sources are employed
~\cite{BerryScheelSandersKnight2004,BerryScheelSandersKnightLaflamme2004}.
Moreover, in order to obtain by means of linear optics a single photon with a
probability $p<1$, which surpasses the efficiencies of the input sources, at
least four such sources are needed. 
A corresponding scheme has been devised in
\cite{BerryScheelSandersKnightLaflamme2004} at the expense of adding an
additional multi-photon component to the output. Combining linear
optics with homodyne detection seems not to improve the efficiency
$p$.  This has been shown at least for the case where only a 
single copy of the state $\rho$ as given in
Eq.~(\ref{imperfect}) is available~\cite{Berry05}.
These results suggest the usage of non-linear optical elements to
improve the efficiency of imperfect single photon sources. 

A possible solution to clean the single photon part in $\rho$ from the
vacuum contribution (to build a ``photon washer'')
might be the usage of a quantum non-demolition measurement as proposed
in \cite{Munro05}. In this article we follow a different road. Our
proposal is based on a Mach-Zehnder interferometer, which contains a
two-photon absorbing medium in one of its arms. 

This article is organized as follows. The main part introduces our
scheme (Sec.~\ref{Scheme}) to generate pure single photons and shows how the
improvement of the efficiency is accomplished
(Sec.~\ref{Calculation}). In Sec.~\ref{Realisations} we sketch
settings where two-photon absorption might occur with sufficiently high
probability to make the scheme feasible. In the appendix we discuss
alternative realizations of  single-photon generation without a 
Mach-Zehnder interferometer. 

\section{The single-photon generator} 
\label{Scheme}

\vspace*{3mm}
\pstex{htb}{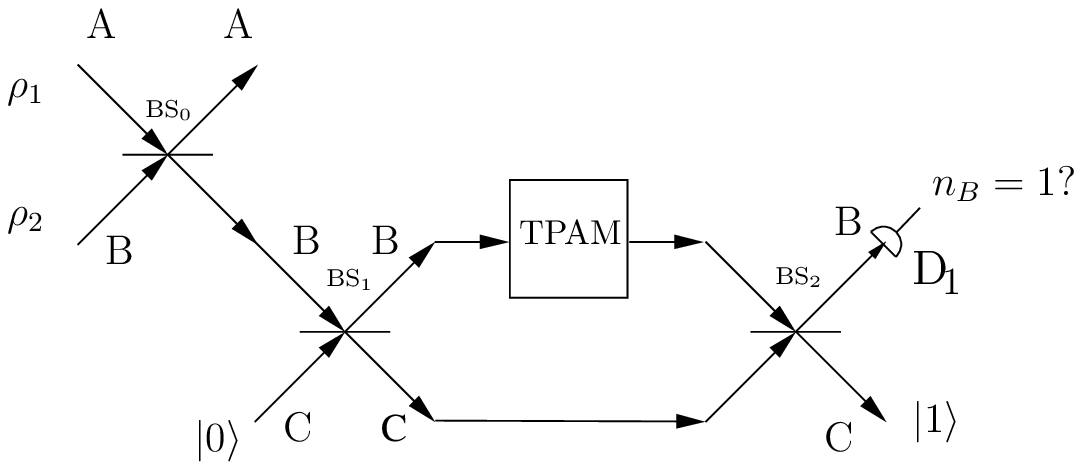}{$\rho_1$ and $\rho_2$ are imperfect single-photon
  states. They are superposed at a  50/50-beam-splitter 
$BS_0$. One of the resulting outputs is ignored (mode $A$) 
whereas the second one (mode $B$) is used as input of a Mach-Zehnder
interferometer, which has  a two-photon-absorbing medium ($TPAM$) in
one of its arms. A successful generation of a single-photon Fock state
$\ket{1}$ is indicated by a click of detector D$_1$ ($n_B=1$).}{Twophoton}

In this section we describe the setup and sketch the underlying physical ideas 
heuristically. The related calculations are then presented in the next
section. 
Our setup is depicted in Fig.~\ref{Twophoton}. The inputs $\rho_1$ and
$\rho_2$ originate from two identical {\em imperfect} single-mode
single-photon sources, with equal efficiency $p$:
$\rho_1=\rho_2=p\ket{1}\bra{1}+(1-p) \ket{0}\bra{0}$,
cf.~Eq.~(\ref{imperfect}). The first step of our 
scheme consists in superposing the two input states 
$\rho_1$ and $\rho_2$ at a  beam splitter  BS$_0$. One 
of the resulting outputs is discarded, whereas the other one is 
processed. The latter is used as input of a Mach-Zehnder
interferometer with a \emph{two-photon absorbing medium} ($TPAM$) 
in one of its arms. The interferometer is then followed 
by a photodetector D$_1$ which is required to click {\em only} 
in the case of a single photon. The detector measures one 
of the interferometer's outputs, its second output contains 
the desired single-photon state $\ket{1}$ whenever a click 
in D$_1$ occurs.

The essential element of our arrangement 
is the two-photon-absorbing medium in one of the arms of 
the interferometer. Its action on the electromagnetic field 
in mode B is assumed to be given by
\begin{eqnarray}
\ket{0^B}\ket{g}&\xrightarrow{TPAM}&
 \ket{0^B}\ket{g}\nonumber\\
\ket{1^B}\ket{g}&\xrightarrow{TPAM}&
 \ket{1^B}\ket{g}\nonumber\\
 \ket{2^B}\ket{g}&\xrightarrow{TPAM}&\alpha\ket{0^B}\ket{e}
 +\beta \ket{2^B}\ket{g}\label{tpam}
\;.
\end{eqnarray}
In transformation~(\ref{tpam})  
the first factor of the state represents the Fock state of  
mode B.  The second factor refers to the {\em collective} quantum
state of the medium. Here 
$\ket{g}$ denotes the \lq\lq ground state'' of the medium 
whereas the \lq\lq excited state'' $\ket{e}$ indicates 
that the medium has absorbed two photons. Please note, 
that it is not necessary in our context to further 
specify the  collective excited state.
Thus, only if two photons propagate through 
the $TPAM$ there is a non-vanishing
probability $|\alpha|^2$ that {\em both} photons are absorbed. 
In case of one or zero photons in mode B no photon absorption 
takes place.. 

For $|\alpha|^2+|\beta|^2=1$ the $TPAM$ acts  
unitarily. To be more general we may also allow the $TPAM$ 
to induce a non-unitary transformation, describing, e.g., a unitary 
process followed by measurements. In the latter 
case $|\alpha|^2+|\beta|^2<1$. Later in this paper we 
will provide two possible realizations of the $TPAM$. 
One of them realizes a unitary $TPAM$ (see Sec.~\ref{OpticalFiber}), 
while the second one implements a non-unitary $TPAM$  (see Sec.~\ref{JF}).

Note that two-photon absorption is possible whenever  
$\alpha\not=0$. In the limiting case $\alpha=0$ 
transformation (\ref{tpam}) describes a {\em non-linear 
phase-shift}, if $\beta/|\beta|\not=1$: the phases of the states $\ket{0}\ket{g}$ and 
$\ket{1}\ket{g}$ are not altered whereas that of state 
$\ket{2}\ket{g}$ is changed.  In general 
both effects, two-photon absorption 
and non-linear phase-shift are present. Nevertheless, in this paper 
we will refer to a medium causing transformation~(\ref{tpam}) as  
a two-photon absorbing medium ($TPAM$)  regardless 
whether the latter or the former effect prevails.  

Beam splitters BS$_1$ and BS$_2$ are chosen such that in 
the absence of the $TPAM$ detector D$_1$
cannot click, 
as a result of destructive interference within the interferometer. 
In this case the photons can leave the setup only in mode $C$. 
The mere existence of the interaction~(\ref{tpam}) leading to entanglement between
the photons and the $TPAM$ destroys, partially, the interference of the
two paths of the  Mach-Zehnder interferometer.    
For two incoming photons the interference is disturbed maximally  in case 
$|\alpha|=1$. In this latter case the $TPAM$ functions as a 
measuring apparatus performing a {\em Null-measurement}. Null-measurements 
are also called, in a slightly misleading way, 
\lq\lq interaction-free measurements''~\cite{Vaidman1993,Vaidman2003}.  
As a consequence, an event becomes possible  
which would be impossible in the absence of the $TPAM$, namely, 
the detection of a single photon at the detector  D$_1$. 
Such a detection is possible 
only if two photons propagate through the interferometer. Thus, when 
detecting a single photon in D$_1$ we know on the one hand that 
two photons have been fed into the setup and on the other hand 
that they have not been absorbed by the $TPAM$. As a consequence, 
if conditioning on detection of a single photon in detector D$_1$, 
we can be sure that there is another single photon leaving 
the setup in mode $C$. A click of the detector D$_1$ {\em announces}  
the presence of a single photon in mode $C$. We thus have 
a {\em heralded} generation of a pure single-photon state   
in mode $C$. 

\section{Theory}
\label{Calculation}

The calculations presented in this section reflect what has 
been said above. Let us start by providing the required 
transformations for a general beam splitter with two input modes $A_1$ and
$A_2$ and two output modes $A'_1$ and $A'_2$
(cf.~Fig.~\ref{Beamsplitter}). 
\pstex{htb}{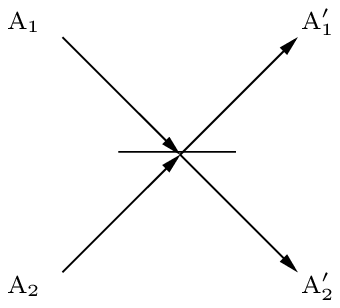}{Beam splitter with two input modes $A_1$ and
$A_2$ and two output modes $A'_1$ and
$A'_2$.}{Beamsplitter}
For the purpose 
of this paper it is convenient to describe the  
action of the beam splitters in the {\em Schr{\"o}dinger picture}. 
A pure input state can be written as $\ket{\Psi}=
f(a^{\dagger}_1,a^{\dagger}_2)\ket{0}$. Here $a^{\dagger}_1$ and
$a^{\dagger}_2$ are the creation operators corresponding to the field
modes  $A_1$ and $A_2$, and $f$ is some functional of them. 
In the same way we can represent the output state by   
$\ket{\Psi'}=f'(a'^{\dagger}_1,a'^{\dagger}_2)\ket{0}$, 
where the creation operators $a'^{\dagger}_1$, $a'^{\dagger}_2$ 
refer to the new field modes  $A'_1$ and $A'_2$ and $f'$ is some 
new functional. In order to obtain the output state $\ket{\Psi'}$ 
from a given input state $\ket{\Psi}$ 
we have simply to perform the following formal replacements 
in $f(a^{\dagger}_1,a^{\dagger}_2)$ (cf.~\cite{Scheel2003}):
\begin{eqnarray}
a_1^\dag &\rightarrow &\tilde{a}^\dag_1= \cos(\theta)a'^\dag_1 + e^{-i\phi}\sin(\theta)a'^\dag_2
\nonumber\\
a_2^\dag &\rightarrow &\tilde{a}^\dag_2= -e^{i\phi}\sin(\theta)a'^\dag_1+\cos(\theta)a'^\dag_2\;,
\label{GeneralBeamsplitterTrafo}
\end{eqnarray}
i.e., $f'(a'^{\dagger}_1,a'^{\dagger}_2)=
f(\tilde{a}^{\dagger}_1,\tilde{a}^{\dagger}_2)$. 
For the sake of simplicity, however, in what follows below, 
we will denote the output field modes by the same letters 
omitting the prime labels, with a convention as already depicted 
in Fig.\ref{Twophoton}.
Note that the replacement transformation (\ref{GeneralBeamsplitterTrafo}) 
should not be confused with the transformation of operators in the 
Heisenberg picture.

The initial input state reads:
\begin{eqnarray}\label{InitialInputState}
\rho_1\otimes\rho_2&=&p^2
\ket{1^A1^B}\bra{1^A1^B}+(1-p)(1-p) \ket{0^A0^B}\bra{0^A0^B}\nonumber\\
&+&(1-p)p \ket{0^A1^B}\bra{0^A1^B}+p(1-p)
\ket{1^A0^B}\bra{1^A0^B}\,\,,
\end{eqnarray}
where the first and the second slot represent the number of photons 
in mode $A$ and mode $B$, respectively. 

\subsection{50/50-beam-splitters}

We will see below that only the two-photon term of
Eq.~(\ref{InitialInputState}) 
is decisive, and that both photons have to be directed into  
mode $B$ by beam splitter BS$_0$. 
Starting with a general transformation 
(\ref{GeneralBeamsplitterTrafo}) for BS$_0$ 
we calculate the density matrix which results 
for mode $B$ after applying BS$_0$ to 
the initial input state $\rho_1\otimes\rho_2$ 
and tracing out the discarded mode $A$: 
\begin{eqnarray}
\rho_1\otimes\rho_2
&\xrightarrow{\mbox{{\small Tr}}_A\,\circ\,\mbox{{\small BS}}_0}\,\rho_B=&
p^2\frac{\sin^2(2\theta_0)}{2}\ket{2}\bra{2}+
\left(p^2\cos^2(2\theta_0)+p(1-p)\right)\ket{1}\bra{1}\nonumber\\
&&+\left( p^2\frac{\sin^2(2\theta_0)}{2}+(1-p)\right)\ket{0}\bra{0}\;.
\end{eqnarray}
Thus, the two-photon part is least diminished for
$\theta_0=\frac{\pi}{4}$. It is therefore favorable to 
use a 50/50-splitter BS$_0$.  In this case  
the density matrix in mode $B$ becomes  
\begin{equation}\label{rho_B_SymmetricBS_0}
\rho_B=
\frac{p^2}{2}\ket{2}\bra{2}+
p(1-p)\ket{1}\bra{1}+ \left(\frac{p^2}{2}-p+1\right)\ket{0}\bra{0}\;.
\end{equation}
In order to make the calculation transparent we assume that also BS$_1$ and
BS$_2$ are 50/50- beam splitters with $\phi=0$, cf.~Fig.~\ref{Twophoton}.
The action of the beam splitters BS$_1$ and BS$_2$ on modes $B$ and
$C$ is then given by:               
\begin{eqnarray}
b^{\dagger}&\rightarrow&\frac{1}{\sqrt{2}}\left(b^{\dagger}+c^{\dagger} \right)\nonumber\\
c^{\dagger}&\rightarrow&\frac{1}{\sqrt{2}}\left(-b^{\dagger}+c^{\dagger}
\right)\;\label{BeamsplitterTrafo}. 
\end{eqnarray}
Later in this section we will report that by employing more 
suitable beam splitters BS$_1$ and  BS$_2$ the 
success probability for a heralded single-photon generation 
can be enhanced.

There are three possible {\em pure} input states that can enter 
the  Mach-Zehnder interferometer in mode $B$. The corresponding 
probabilities appear in Eq.~(\ref{rho_B_SymmetricBS_0}). 
We discuss the three  alternative cases separately. 
The case of zero photons which occurs with probability
$\mbox{prob}(0)=\left(\frac{p^2}{2}-p+1\right)$ 
trivially cannot be related with a click 
at the second single-photon detector D$_1$.

Let us now consider the case that  
exactly one photon is fed into the setup. It can come either from the
input $\rho_1$ in mode A  or from  input $\rho_2$ in mode B. The over-all 
probability for a single photon being in mode $B$ amounts to  
$\mbox{prob}(1)=p(1-p)$. The photon interferes 
with the vacuum (in mode $C$) at the  beam-splitter ${BS_1}$. 
In the following the first slot of the kets refers to mode 
$B$ and the second slot to mode $C$. The input state for 
the  Mach-Zehnder interferometer is thus given by
$\ket{1^B0^C}=b^\dag\ket{0^B0^C}$. 
The calculations yield with Eqs.~(\ref{tpam}) and (\ref{BeamsplitterTrafo}):
\begin{eqnarray}
\ket{1^B0^C}=b^\dag\ket{0^B0^C}&\xrightarrow{BS_1}
& \frac{1}{\sqrt{2}}(b^\dag+c^\dag)\ket{0^B0^C}\nonumber
\\ &\xrightarrow{TPAM}&
\frac{1}{\sqrt{2}}(b^\dag+c^\dag)\ket{0^B0^C}\nonumber\\ 
&\xrightarrow{BS_2}& 
\frac{1}{\sqrt{2}}\left(\frac{1}{\sqrt{2}}(b^\dag+c^\dag)+
\frac{1}{\sqrt{2}}(-b^\dag+c^\dag)\right)\ket{0^B0^C}\nonumber\\
&\xrightarrow{\dm{1^B}{1^B}\otimes\openone^C}& 0\;.
\end{eqnarray}
In the last line  $\dm{1^B}{1^B}\otimes\openone^C$  
represents a projective selective 
measurement on mode $B$, the result being 
a detection of a single photon. 
Thus, just one photon entering the setup in mode $A$ or $B$ cannot cause  a click 
in detector  D$_1$. 

What happens in the case in which two photons are fed into the setup,
one in mode A and one in mode B? In the calculation below we will need 
to take into account also the state of the medium as soon as the 
measurement-like interaction inside the TPAM is applied. 
Furthermore, we will discard the mode which is detected. 
Again, the first slot of the kets refers to mode 
$B$ and the second slot to mode $C$, whereas the 
third slot represents the state of the medium.  
The input state for the  Mach-Zehnder interferometer is now given by
$\ket{2^B0^C}=\frac{1}{\sqrt{2}}(b^\dag)^2\ket{0^B0^C}$. Applying 
50/50-beam-splitters BS$_1$ and BS$_2$, the $TPAM$, and 
conditioning on the detection of a single photon by detector
D$_1$, yield: 
\begin{eqnarray}
\ket{2^B0^C}=\frac{1}{\sqrt{2}}(b^\dag)^2\ket{0^B0^C}
&\xrightarrow{BS_1}& \frac{1}{2\sqrt{2}}
(b^\dag+c^\dag)^2\ket{0^B0^C}\nonumber\\&&=
\frac{1}{2\sqrt{2}} (\sqrt{2}\ket{2^B0^C}+\sqrt{2}\ket{0^B2^C}+2\ket{1^B1^C})\nonumber
\\
&\xrightarrow{TPAM}&\frac{1}{2\sqrt{2}}
\Big(\phantom{\Big)}\sqrt{2}(\alpha\ket{0^B0^Ce}+\beta\ket{2^B0^Cg})+\nonumber\\&&
+\phantom{\Big(}\sqrt{2}\ket{0^B2^Cg}+2\ket{1^B1^Cg}\Big)\nonumber\\&&=
\frac{1}{2\sqrt{2}}\Big(\phantom{\Big)} \sqrt{2}\alpha\ket{0^B0^Ce}+
\beta (b^\dag)^2\ket{0^B0^Cg}+\nonumber\\&& +\phantom{\Big(}
(c^\dag)^2\ket{0^B0^Cg}+2b^\dag c^\dag\ket{0^B0^Cg}\Big)
\nonumber \\
&\xrightarrow{BS_2}&\frac{1}{2\sqrt{2}}\Big(\beta
 \frac12(b^\dag+c^\dag)^2\ket{0^B0^Cg}+\frac12(-b^\dag+c^\dag)^2\ket{0^B0^Cg}\phantom{\Big)}
\nonumber\\&&+\phantom{\Big(}
(b^\dag+c^\dag)(-b^\dag+c^\dag)\ket{0^B0^Cg}+\sqrt{2}\alpha\ket{0^B0^Ce}\Big)\nonumber
\\&&= 
\frac{1}{2\sqrt{2}}\bigg(\Big(\frac{1}{\sqrt{2}}
(\beta+1)-\sqrt{2}\Big)\ket{2^B0^Cg}+(\beta-1)\ket{1^B1^Cg}
\phantom{\Big)}\nonumber\\&&\phantom{\Big(}+
\Big(\frac{1}{\sqrt{2}}(\beta+1)+\sqrt{2}\Big)\ket{0^B2^Cg}+
\sqrt{2}\alpha\ket{0^B0^Ce}\bigg)\nonumber\\
&\xrightarrow{\dm{1}{1}\otimes\openone
\otimes\openone}&
\frac{1}{2\sqrt{2}} (\beta-1) \ket{1^Cg}\;.
\end{eqnarray}
The final projective measurement by detector D$_1$ 
breaks the entanglement between the  $TPAM$ and the photons. 
The $TPAM$ is transferred in a non-local 
way to the ground state $\ket{g}$.  

The above calculation shows that, whenever two photons are fed into the setup 
there is a non-vanishing probability for a click event 
at the final detector D$_1$. Moreover, on the condition of this event one 
can be sure that another single photon is leaving our device in mode
$C$. A detection of a single photon by detector D$_1$ 
{\em guarantees} a \emph{perfect} preparation of a single photon Fock 
state $\ket{1}$ in mode $C$.  The probability for a click in detector D$_1$,  
given that two photons are injected into the Mach-Zehnder
interferometer in mode $B$, amounts to   
$\frac{1}{8}|1-\beta|^2$, as can be inferred from the norm 
of the resulting (unnormalized) final quantum state. 
With the probability for two photons 
entering the Mach-Zehnder interferometer in mode $B$ being  
$\mbox{prob}(2)=p^2/2$ (cf.~Eq.(\ref{rho_B_SymmetricBS_0})), 
the over-all \emph{probability 
for success} thus amounts to 
\begin{equation}
P^{50/50}_{\mbox{\scriptsize
    s}}=\frac{1}{16}|1-\beta|^2p^2 \;.
\end{equation}

\subsection{The general case}
The success 
probability $P_{\mbox{\scriptsize  s}}$ 
can be increased by employing more suitable beam splitters
BS$_1$ and BS$_2$. This follows from an analysis of the scheme using    
general beam-splitter transformations~(\ref{GeneralBeamsplitterTrafo}) 
for BS$_1$ and BS$_2$. We omit the details of the calculations. 

The requirement that a single photon entering the
setup either in mode $A$ or in mode $B$ must not cause  a click 
in the detector D$_1$ restricts the possible values of the 
beam-splitter parameters $\phi_1,\phi_2,\theta_1,\theta_2$. 
This requirement firstly implies that $(\phi_1-\phi_2)=\nu\pi$ 
with $\nu\in\mathbb{Z}$. 
The conditions for $\theta_1$ and $\theta_2$ depend on whether 
$(\phi_1-\phi_2)=\nu(2\pi)$ or $(\phi_1-\phi_2)=(2\nu+1)\pi$, with $\nu\in\mathbb{Z}$. 
In the first case we must have  $(\theta_1+\theta_2)=\pm \frac{\pi}{2}$, 
while in the second case we must choose  $\theta_1$ and $\theta_2$ such that 
 $(\theta_1-\theta_2)=\pm \frac{\pi}{2}$. Taking into account these 
conditions we obtain the following: If we choose $(\phi_1-\phi_2)=\nu(2\pi)$ and 
$(\theta_1+\theta_2)=+\frac{\pi}{2}$ or $(\phi_1-\phi_2)=(2\nu+1)\pi$ and 
$(\theta_1-\theta_2)=+\frac{\pi}{2}$, the calculation yields  
\begin{equation}
P_{\mbox{\scriptsize s}}(\beta,\theta_1,\theta_2)=p^2|1-\beta|^2
\cos^6(\theta_1)\sin^2(\theta_1)\;,  \label{SuccessFirstCase}
\end{equation}
whereas the choices  $(\phi_1-\phi_2)=\nu(2\pi)$ and 
$(\theta_1+\theta_2)=-\frac{\pi}{2}$ or $(\phi_1-\phi_2)=(2\nu+1)\pi$ and 
$(\theta_1-\theta_2)=-\frac{\pi}{2}$ lead to  
\begin{equation}
P_{\mbox{\scriptsize s}}(\beta,\theta_1,\theta_2)=p^2|1-\beta|^2
\sin^6(\theta_1)\cos^2(\theta_1)\;.   \label{SuccessSecondCase}
\end{equation}
Please note, that the optimal reflectivity of BS$_1$ and BS$_2$ does not depend
on the parameters characterizing the $TPAM$.

The success probability becomes maximal for
$\theta_1^{\mbox{\scriptsize max}}=30^\circ,150^\circ,  
210^\circ,330^\circ$ in the first case~(\ref{SuccessFirstCase}) and 
for  $\theta_1^{\mbox{\scriptsize max}}=60^\circ,120^\circ,240^\circ,300^\circ$ in the second 
case~(\ref{SuccessSecondCase}). The most suitable  beam splitters thus turn out to be a
$BS_1$ with  reflectivity of $3/4$ and a BS$_2$ with  reflectivity 
of $1/4$, or vice versa. In addition the phase condition $(\phi_1-\phi_2)=\nu
\pi$ has to be fulfilled. The maximal success probability that 
can be reached by variation of the beam-splitter parameters, is 
\begin{equation}
P^{\mbox{\scriptsize max}}_{\mbox{\scriptsize s}}(\beta)\equiv
P_{\mbox{\scriptsize s}}(\beta,\theta^{\mbox{\scriptsize max}}_1,
\theta^{\mbox{\scriptsize max}}_2)=
\frac{27}{256}p^2|1-\beta|^2\approx 0.1055\times p^2|1-\beta|^2 \;.  
\end{equation}
It still depends on $\beta$, which characterizes the 
action of the $TPAM$, cf.\ Eq.~\Ref{tpam}. 
$P^{\mbox{\scriptsize max}}_{\mbox{\scriptsize
    s}}(\beta)$ becomes zero, if there is no $TPAM$ in one of the 
interferometer arms, i.e.,\ in case $\beta=1$.  
$P^{\mbox{\scriptsize max}}_{\mbox{\scriptsize
    s}}(\beta)$ becomes maximal for $\beta=-1$, which means no
two-photon absorption but just a non-linear phase shift. 
In the latter case a success probability 
$P^{\mbox{\scriptsize max}}_{\mbox{\scriptsize s}}(\beta=-1)\approx 
0.4219\times p^2$ is attained. 
Note, that to alter the phase of $\beta$ by introducing a global phase factor
in the third transformation rule of (\ref{tpam}) would entail the
introduction of the same phase factor in the first two transformation
rules of  (\ref{tpam}). The calculations show that  probability
$P^{\mbox{\scriptsize max}}_{\mbox{\scriptsize
    s}}(\beta)$ is left invariant, as it is expected,  under
such a global phase change. In the next section we 
consider realizations of $TPAM$s 
with real positive values for $\beta$. In this case 
$P^{\mbox{\scriptsize max}}_{\mbox{\scriptsize
    s}}(\beta)$ increases with growing two-photon 
absorption rate. It assumes its maximal value 
$P^{\mbox{\scriptsize max}}_{\mbox{\scriptsize
    s}}=\frac{27}{256}p^2$ for 
$|\alpha|=1$ or $\beta=0$, respectively.

\vspace*{3mm}
\pstex{uh!}{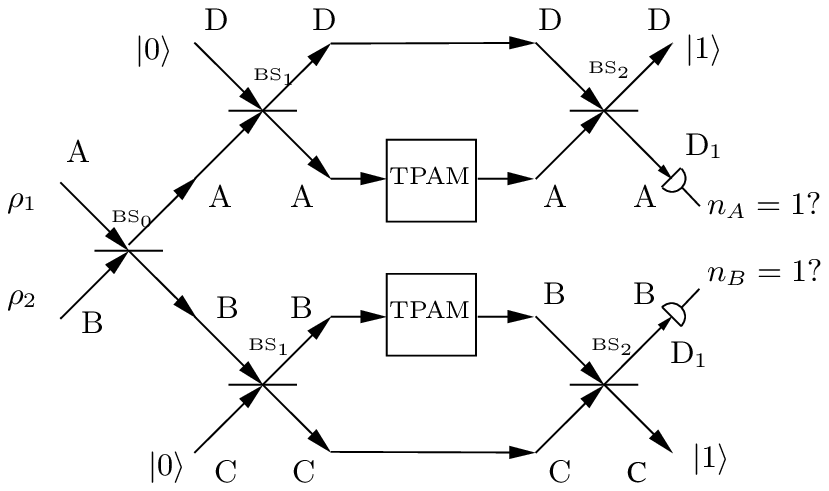}{Enhancement of the success probability 
{\em by a factor of} 2 by processing {\em both} outputs of BS$_0$ 
using identical interferometers with a $TPAM$ and final detectors
D$_1$ --- instead of discarding one of them as done in the 
original setup (cf.~ Fig.\ \ref{Twophoton}).}{DoubleTPAM}

A further enhancement of the success probability {\em by a factor of} 2 can be achieved by processing
the second output of BS$_0$ in the same way as the first one
(cf.~Fig.~\ref{DoubleTPAM}),  instead of
discarding it. The second output is directed into an
identical interferometer with a $TPAM$ and a final detector D$_1$. 
Provided that two ingoing photons take the upper way
this alternative heralds a pure single photon with the same probability as in
the lower case. Since both photons choose either the upper or the
lower path with equal probability --- due to
the Hong-Ou-Mandel effect --- the two alternatives are exclusive. 
Therefore the corresponding success probabilities for heralding a
photon are additive. 
As a result, the over-all maximum success probability that can 
in principle be achieved  with such an extension of our scheme amounts
to 
$2\times P^{\mbox{\scriptsize max}}_{\mbox{\scriptsize
    s}}(\beta=-1)=2\times 4\times\frac{27}{256} p^2 = 0.84375\times
p^2$. This is  remarkably high as compared to the upper bound $p^2$ 
of the probability for a heralded generation of a single photon from two identical sources 
with efficiency $p$.

\section{Realization}
\label{Realisations}

We believe that the scheme proposed above is interesting in its own
right.  Whether our method to generate a single photon from two
imperfect sources can already be applied in a  laboratory 
depends on the availability of a
setup which realizes state transformation (\ref{tpam}), such that the
success probability $P_{\mbox{\scriptsize s}}\propto |1-\beta|^2$ is sufficiently high.   
In the following we are going to suggest two 
candidates for the implementation of state transformation (\ref{tpam}).
We will refer to each of them as two-photon absorbing medium (TPAM), even
though they rather resemble machines built out of several units. 
Both TPAMs use special settings to amplify 
two-photon absorption, which normally is a much weaker effect than 
absorption or scattering of a single photon. These settings stem from 
different contexts. We propose modifications which make them suitable
for our purpose.

\subsection{TPAM realized by a thin hollow optical fiber
  containing three-level atoms} 
\label{OpticalFiber}

A TPAM is realized whenever the medium can absorb two photons, whereas
just one photon is not absorbed.  
A good model for understanding two-photon absorption is given 
by  three-level atoms (cf.~\cite{FransonJacobsPittman2004}). 
Fig.~\ref{TPA2} illustrates how two-photon absorption 
can take place, while single-photon
absorption is impossible due to detuning.

\vspace*{3mm}
\pstex{bh}{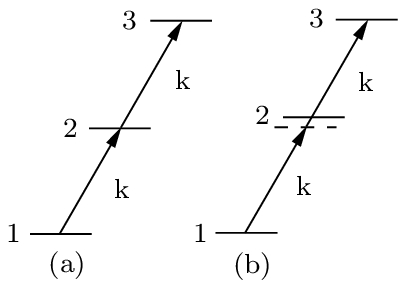}{Illustration of two-photon absorption by means 
of {\em three-level atoms} (cf.~\cite{FransonJacobsPittman2004}). 
The atomic levels are labeled by 1, 2 and 3 
while the photons are indicated by their wave numbers $k$. 
In (a) a single photon can be absorbed. This is avoided by detuning
(b). 
}{TPA2}

Two-photon absorption is a nonlinear effect and is therefore commonly 
expected to be small for intensities corresponding to a few photons only. 
A way to amplify this nonlinear process is to insert three-level atoms
with appropriate detuning  
into a thin hollow optical fiber. Such a proposal has recently been 
discussed by Franson~et\ al.\ in~\cite{FransonJacobsPittman2004} in
the context of optical realizations of quantum gates.  
There it is argued that the confinement of a single-photon wave
packet within a very narrow optical fiber involves concentration of
the photon energy into a very small volume. The confinement can thus produce relatively
large electric fields which in turn entail the possibility of large
nonlinearities, including two-photon absorption. According to 
estimates provided in~\cite{FransonJacobsPittman2004} it is 
feasible to achieve two-photon absorption rates in optical fibers 
that correspond to a two-photon absorption length on the 
order of $5m$. Given that such values can in fact be 
achieved, we believe that $|\alpha|$-values (cf.~Eqs.~\ref{tpam}) 
close to $1$ are feasible. Since in this realization  of the  
$TPAM$ no measurements take place, the transformation~(\ref{tpam}) 
is unitary and therefore we have $|\beta|^2=1-|\alpha|^2\approx 0$ 
for high two-photon absorption rates, which involves a success probability 
$P^{\mbox{\scriptsize max}}_{\mbox{\scriptsize s}}(\beta\approx 0)\approx
\frac{27}{256}p^2\approx 0.1055\times p^2$.

Certain losses need to be overcome that might cause failure events 
of such a realization of the single-photon generator. The main
technical challenge seems to consist in suppressing the single-photon 
scattering which, under most conditions, is expected to prevail 
the two-photon absorption. 
However, scattering of a single photon into another mode 
is suppressed due to the fact that only 
the propagation of a certain mode is supported 
within a {\em single-mode} optical fiber.  
The mere fact that quantum key distribution is 
possible, which relies on the feasibility to 
transmit single photons through 
several kilometers of optical fibers,  
is a strong evidence that losses due to 
single-photon scattering can be made 
negligible.

Additional losses might be caused when coupling the photons 
into the optical fiber. 
This problem can be circumvented by using imperfect single photon sources that 
provide photons {\em inside} a single-mode optical fiber {\em at the
  outset}. 
Such imperfect single photon sources have been reported on, e.g., 
in~\cite{Fasel04,Pittman04}. Moreover, it is experimentally feasible 
to realize beam splitters and phase shifters by means of optical fibers
(see e.g.\ \cite{Fasel04, Pittman04}). Thus, 
our setup of Sec.~\ref{Scheme} can be built in such 
a way that all  processes apart from the final detection 
take place within optical fibers. 
We believe that all these problems can be tackled so as to 
make our proposal of Sec.~\ref{Scheme} and \ref{Calculation} promising.

\subsection{TPAM using resonant nonlinear optics and time selection}
\label{JF}

In this section we introduce another method to realize
a two-photon absorbing medium (TPAM), which we refer to as
``time selection''. We first outline the
principle. Then we describe a setup with non-linear optics proposed by
Johnsson and Fleischhauer~\cite{JohnssonFleischhauer2003} and
modify it to realize a TPAM based on time selection.

The $TPAM$ in our scheme  has the task to only absorb a
pair of photons while being transparent for a single photon. The
obvious way to tackle this task is by means of energy selection. That
is to choose a medium whose transition energy is on resonance with
the energy of two photons but not with the energy of a single
photon. The required two photon process is a second order effect and
has to be enhanced by a special setting to occur with a reasonable
probability. The setup described in the previous section goes along
these lines by confining the electromagnetic field in an optical
fiber. 
We now turn to a different scheme. 

An alternative to energy selection might be ``time
selection'', which will be explained in the following. 
Let us assume that the state of light in the medium traverses a
cycle with a period which depends on the initial number of photons.
The length of the medium can then be adjusted to the time $T_1$ it takes for
a single photon state to reoccur 
\begin{equation}\label{time1}
\ket{1}\ket{g}\xrightarrow{U(T_1)} \ket{1}\ket{g}\,.
\end{equation}
Here $\ket{g}$
represents the initial state (ground state) of the medium.
Two photons run through a different cycle with period 
$T_2\not=T_1$ and emerge after time $T_1$
from the medium in an entangled state 
\begin{equation}\label{time2}
\ket{2}\ket{g}\xrightarrow{U(T_1)}
\alpha_0\ket{0}\ket{e_2}+\alpha_1\ket{1}\ket{e_1} +\beta
\ket{2}\ket{g}\,,
\end{equation}
where $\ket{e_1}$ and $\ket{e_2}$ are excited states of the medium
corresponding to one and two absorbed photons, respectively.
If zero photons enter the
medium, then also zero photons will emerge form it:
\begin{equation}\label{time3}
\ket{0}\ket{g}\xrightarrow{U(T_1)} \ket{0}\ket{g}\,.
\end{equation}

Eqs.\ (\ref{time1}), (\ref{time2}) and  (\ref{time3}) already resemble
the state transformation (\ref{tpam}) required for the TPAM apart from
the one photon term in Eq.\  (\ref{time2}).  This term has to be
eliminated because a one photon contribution after the TPAM could
trigger detector $D_1$ leaving vacuum in the output of our single
photon generator (cf.\ Fig.\ \ref{Twophoton}). One method to remove
the one photon contribution in Eq.\  (\ref{time2}) would be a
conditional measurement projecting on a subspace
orthogonal to state $\ket{e_1}$. For this purpose, however, the
corresponding degree of freedom of the medium has to be accessible by
measurement. 
 
Johnsson and Fleischhauer (JF) \cite{JohnssonFleischhauer2003} 
proposed a setup using resonant four-wave mixing
with two pump fields $\Omega_1$, $ \Omega_2$ and two generated fields
$E_1$, $E_2$, where the energy cycles between
the pump- and the generated fields. In their
scheme  the
light fields interact via a vapor of five-level atoms (see Fig.\
\ref{fivelevel}) which can be understood as  modified double $\Lambda
$-systems. The additional level $\ket{3}$ serves to compensate
non-linear phase shifts of the light fields, which would prevent 
energy cycling with unit efficiency. The modification also reduces the
period of the cycle as compared to a generic double-$\Lambda$
system \cite{JohnssonFleischhauer2002}. Initially the atom is in state
$\ket{1}$ and the first pump field $\Omega_1$ contains $n$
photons. The light fields $E_1$ and $E_2$ are initially not
excited but generated by the interaction of light and matter. The
second pump field $\Omega_2$ consists of a strong coherent cw input
which enhances the cycling. All fields propagate in the same direction. 
Choosing a driving field $\Omega_1$ in
resonance with the $\ket{2}-\ket{5}$ transition  and a driving field
$\Omega_2$ with detuning $\mp\Delta$ with respect to the
$\ket{1}-\ket{3},\ket{4}$ transition, minimizes losses due to single
photon absorption. It can be shown that the fields $E_1$ and $E_2$ are
then generated precisely with frequencies, such that there are
two-photon resonances from  the pair $\Omega_1$, $E_1$ and from the
pair $\Omega_2$, $E_2$ corresponding to $\ket{1}-\ket{2}$ transitions. 
This results in an overall four-photon resonance. 

\pstex{h!}{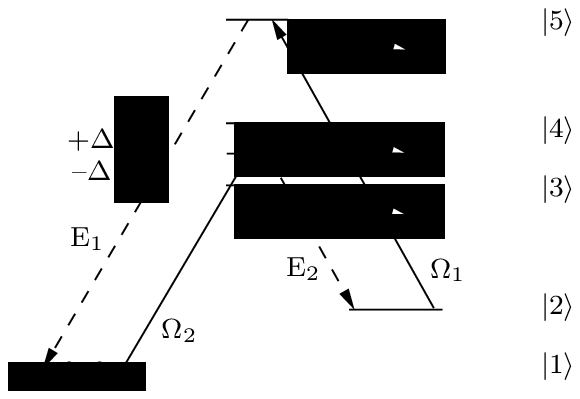}{Four-wave mixing scheme in a modified
  double-$\Lambda$ 
  system according to Johnsson and Fleisch\-hauer~\cite{JohnssonFleischhauer2003}. In order to
  avoid non-linear phase shifts, $d_{42}d_{41}d_{32}d_{31}<0$, where
  $d_{ij}= \bra{i}d\ket{j}$ is the matrix element of the dipole moment
  corresponding to the $\ket{i}-\ket{j}$ transition. 
  $\gamma_1,\gamma_2,\gamma_3$ indicate spontaneous decay rates of the 
  corresponding energy levels. The level scheme
  is realized in the hyperfine structure of atoms \cite{JohnssonFleischhauer2002}.} {fivelevel}

According to the results of JF \cite{JohnssonFleischhauer2003}
the evolution of the initial state $\ket{\psi(0)} =
\ket{1,0,0}$ is given by
\begin{equation}
  \ket{\psi(t)}= \cos(\kappa|\Omega_2|ct)\ket{1,0,0}-
  i\sin((\kappa|\Omega_2|ct)\ket{0,1,1}\,, \label{Rabi}
\end{equation}
where $\ket{l,m,n}$ represents the Fock state with $l,m$ and $n$ photons in
the modes corresponding to the fields $\Omega_1$, $E_1$ and $E_2$,
respectively. The coherent cw field $\Omega_2$ can be treated
classically. Here $\kappa:= 3N\lambda^2\gamma/(8\pi\Delta)$ and $N$ is
the atomic number density, $\lambda$ and $\gamma$ are  some typical
wavelength and radiative decay rate, respectively  (cf.~\cite{JohnssonFleischhauer2003}). 
Eq.\ (\ref{Rabi}) indicates an oscillation between 
states with one photon in $\Omega_1$ and one photon in $E_1$ and $E_2$ each.  
Following the scheme of JF we restrict the
length of the nonlinear medium to a multiple $M$ of
$L_0=\pi/(\kappa|\Omega_2|)$. A medium with such a length will be transparent
for a single photon in $\Omega_1$:
\begin{equation}
\ket{1,0,0}\xrightarrow{U(t=M L_0/c)} (-1)^M\ket{1,0,0}\;,
\end{equation}
except for a phase shift for odd $M$, which can be compensated for by an
additional phase shifter.
This corresponds to the method of ``time
selection'' mentioned above, cf.\ Eq.\ (\ref{time1}). But here
the role of the state of the medium $\ket{g}$ in Eq.\ (\ref{time1})
is played by the state of the radiation fields $E_1$, $E_2$. This is
advantageous since this state is accessible by measurement.   

Starting with two photons in $\Omega_1$, i.e.\  $\ket{\psi(0)} =
\ket{2,0,0}$, one obtains the following state after time $t=ML_0/c$:
  
\begin{equation}
\label{Rabi2}
\ket{2,0,0}\xrightarrow{U(t=M L_0/c)}\alpha_0\ket{0,2,2}+\alpha_1\ket{1,1,1}
+ \beta\ket{2,0,0}\,.
\end{equation}
The values of $\alpha_0$, $\alpha_1$ and $\beta$ as functions of the
interaction time are given by (cf.\ ~\cite{JohnssonFleischhauer2003})
\begin{eqnarray}
&\alpha_0(t) =   
-\dfrac{2\sqrt{2}}{3}\dfrac{\Omega_2}{\Omega_2^*}\sin^2\left(\frac{\varphi(t)}
{2}\right) \quad &\alpha_1(t)=
-\frac{i}{\sqrt{3}}\sqrt{\frac{\Omega_2}{\Omega_2^*}} \sin\left(
\varphi(t)\right) \nonumber\\
&\beta(t) = \dfrac{1}{3}\left[2+\cos\left(\varphi(t)\right)\right]\quad
&\mbox{with} \quad \varphi(t):=\sqrt{\frac{3}{2}}\kappa|\Omega_2|c t\,.
\label{Rcoeff}
\end{eqnarray}

Now the second term in Eq.\ (\ref{Rabi2}) represents a one photon
contribution and has to be eliminated to yield a state transformation
of form (\ref{tpam}), which realizes the TPAM. This can be done by
detecting the photons of the generated fields $E_1$ and $E_2$. The
proposal of JF also contains such a detection.
They assume a beam splitter which transmits the pump fields $\Omega_i$ and
reflects  the generated fields $E_i$ ($\,i=1,2$), followed by a detector
for the $E_i$ (cf. BS$_4$ of Fig.\ref{tpam2}). Such a beam
splitter can, e.g., be realized by choosing orthogonal polarizations for the
$\Omega_i$ and  the $E_i$ and using a polarizing beam splitter.
Alternatively a system
of dichroic beam splitters could be employed, which transmit or reflect
light depending on its wavelength. 

\pstex{h!}{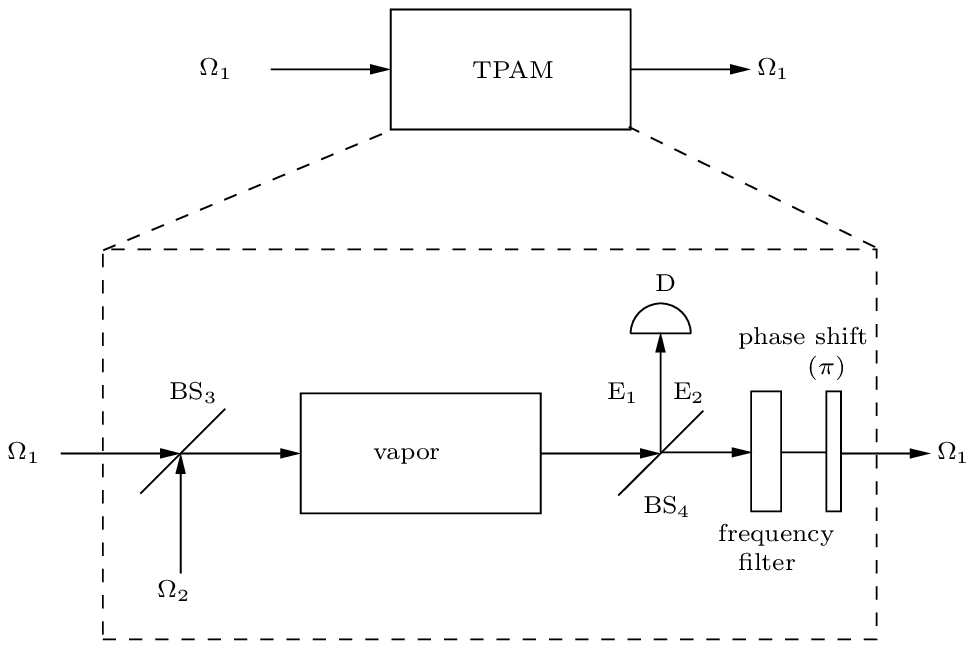}{Implementation of the four-wave mixing
  scheme. $\Omega_1$ is identical with mode B of the
Mach-Zehnder interferometer in  Fig.~\ref{Twophoton} }{tpam2}

At this point we deviate from the
proposal of JF. They consider a sequence
of nonlinear media with length $L_0$ and detectors which makes  the
state of the radiation field converge to  a mixture
of $\ket{1,0,0}$ and vacuum with probability one. Our objective, however, is to
eliminate the vacuum contribution and obtain a single-photon state. 
For this purpose we suggest to condition on the detection of
zero photons in $E_1$ and $E_2$.  This leads to the following state
transformation   
\begin{equation}\label{Rabi2p}
 \ket{2,0,0}\xrightarrow{\left(\openone\otimes\dm{0}{0}\otimes\dm{0}{0}\right)\,
 U(t=L_0/c)} \beta\ket{2,0,0}
 \end{equation}
with $\beta=0.4130$. With this result a special instance
of the desired transformation (\ref{tpam}) is
established and the TPAM comprising the vapor of five-level atoms,
the driving  field $\Omega_2$ and a detector $D$ can be implemented in
one arm of the Mach-Zehnder interferometer (cf. Fig.~\ref{Twophoton})
as shown in Fig.~\ref{tpam2}. The coherent cw field $\Omega_2$
generated by a laser  is superposed with $\Omega_1$ (Mode B of the
Mach-Zehnder interferometer in  Fig.~\ref{Twophoton}) by means of a
dichroic beam splitter. At the end of the TPAM the cw-field $\Omega_2$  
has to be filtered out by means of a frequency filter. The value of $\beta$ in        
transformation (\ref{Rabi2p}) leads to a success probability of 
$P_{\mbox{\scriptsize s}}\approx 0.0363\times p^2$ for the generation of a single photon.
Please note, that due to the fact that $\varphi(t=ML_0/c)=
M\pi \sqrt{3/2}$ in $\beta$ as given by (\ref{Rcoeff}) is an irrational multiple
of $\pi$, any value of $\varphi\in [0,2\pi]$ is assumed with
arbitrary accuracy when  choosing $M$, i.e., the length of the medium,
appropriately. The success probability $P_{\mbox{\scriptsize s}}$ can in
principle be increased up to the value $P_{\mbox{\scriptsize
    s}}^{\mbox{\scriptsize max}}\approx 0.0469\times p^2$ for $\beta=\frac{1}{3}$. Already for a medium of length
$4L_0$ a success probability of $P_{\mbox{\scriptsize s}}\approx 0.0446 \times
p^2$ is accomplished.

The JF setup can thus be modified and employed as TPAM within
  our scheme  in order to generate single-photon Fock states from two
  outputs of imperfect single photon sources. But this is not the only
  way to accomplish this task. In the appendix we sketch two
  alternative schemes without Mach-Zehnder interferometer to remove 
  the vacuum contribution, which is still present in the output of original JF
  setup.

\section{Conclusion}
The scheme to generate single photons as proposed in sections 1 and 2
is clear cut and simple from a conceptual point of view. It also allows to
generate a single photon heralded by a click of detector $D_1$  with
an efficiency of $100\%$ from two single-photon sources with any
efficiency $1>p>0$. In principle, our proposal allows for a success 
probability $P_{\mbox{\scriptsize s}}\approx 0.84\times p^2$. 
However, whether it is experimentally applicable,
e.g.\ in a quantum computation or a quantum cryptography protocol
which relies on single photons, depends essentially on the existence of a
medium or a machine which implements the state transformation given
in Eq.\ (\ref{tpam}) and gives rise to a  sufficiently high success
probability $P_{\mbox{\scriptsize s}}\propto |1-\beta|^2$. 
We propose two candidates to realize this transformation. 

The first employs three-level atoms which can
only absorb a pair of photons because of energy conservation. They are
enclosed in an optical fiber for two reasons. Firstly, confining
the electromagnetic field in the fiber increases the energy density
 and thus enhances non-linear effects such as two-photon
absorption. Secondly the optical fiber may carry only certain 
modes and hence suppresses scattering of single photons. A fiber of
approximately. $5 m$  length yields a sufficiently high probability for two photon absorption.  

The second candidate employs resonant four-wave mixing in a vapor of
five-level atoms with  a strong coherent cw driving field $\Omega_2$
in addition
to the incident light field $\Omega_1$ from the imperfect sources. 
Due to the four-wave mixing two additional light fields $E_1$ and
$E_2$ are generated and the energy cycles between the four fields. 
The period depends on the number of photons $\Omega_1$ initially contains.   
The length of the medium is adjusted such that a single photon in
$\Omega_1$ leaves the atomic vapor unchanged while two photons are
entangled with the generated fields $E_1$ and $E_2$. A conditioned
measurement of the number of photons in $E_1$ and $E_2$ realizes state
transformation (\ref{tpam}) leading to reasonably high success
probabilities $P_{\mbox{\scriptsize s}}$.

\vspace*{3mm}
\noindent{\bf Acknowledgment:} We would like to thank Alex Lvovsky, Lajos Diosi,
Christian Kasztelan, Rudolf Bratschitsch, Florian Adler, and Matthias Kahl
for helpful discussions. This work was supported by the \lq\lq Center
for Applied Photonics'' (CAP)  at the University of Konstanz.

\section*{\large Appendix: Single-photon generation  
without Mach-Zehnder interferometer}
We found two alternative modifications of the setup of Fleischhauer
and Johnsson (see Sec.\
\ref{JF})   to
generate a pure single-photon state out of two noisy photon states
$\rho_1=\rho_2=p\ket{1}\bra{1}+(1-p) \ket{0}\bra{0}$. Both schemes
deviate from the mechanism described in Sec.\ \ref{Scheme} and
\ref{Calculation}, since they operate without a Mach-Zehnder
interferometer. They are depicted in Figs.\ \ref{FigA} and \ref{FigB}.
The alternative schemes use as central unit the implementation of the
TPAM by four-wave mixing which is described above (cp.\ Fig.~
\ref{tpam2}). The input field $\Omega_1$ of the TPAM consists of  the  mixture
(\ref{rho_B_SymmetricBS_0}) gained from superposing 
state $\rho_1$ and $\rho_2$ at a
symmetric beam splitter BS$_0$.   

\pstex{h!}{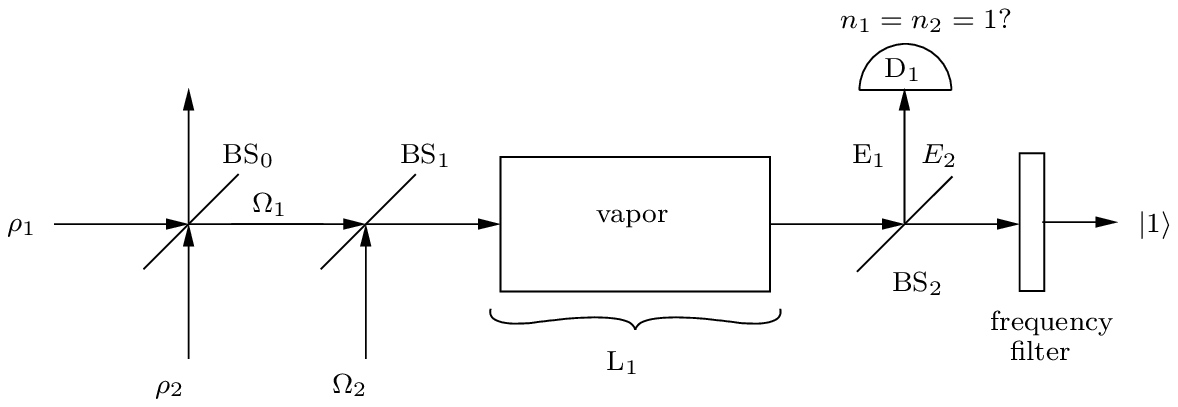}{Alternative setup based on the scheme of JF with
  nonlinear medium (vapor) of length $L_1=2L_0$. Detector D$_1$ 
is conditioned on 1 photon in $E_1$ and $E_2$ each. The success 
probability amounts to $P_{\mbox{\scriptsize s}}\approx 0.1620 \times p^2$}{FigA}

The first alternative uses a vapor of length
$L_1=2L_0$ with $L_0=\frac{\pi}{\kappa|\Omega_2|}$ corresponding to a full cycle of
a single  incident photon in $\Omega_1$, cf.\ Eq.\ (\ref{Rabi}): 
\begin{align}
&\ket{0,0,0}\xrightarrow{U(t=L_1/c)} \ket{0,0,0}\;,\nonumber\\
&\ket{1,0,0}\xrightarrow{U(t=L_1/c)} \ket{1,0,0}\;,\nonumber\\
&\ket{2,0,0}\xrightarrow{U(t=L_1/c)} \alpha_0^{(L_1)}\ket{0,2,2}+\alpha_1^{(L_1)}\ket{1,1,1}
+ \beta^{(L_1)}\ket{2,0,0}\,.
\end{align}
Conditioning on the detection of one photon in $E_1$ and $E_2$,
respectively, eliminates the contributions from a single incident
photon and vacuum in $\Omega_1$ while it selects the single-photon contribution  from
two incident photons. Thus the over-all state change of the scheme
depicted in Fig.~\ref{FigA} amounts to 
\begin{equation}
  \rho_1\otimes\rho_2\rightarrow \frac{p^2 |\alpha_1^{(L_1)}|^2}{2}\ket{1}\bra{1}
\end{equation}
This process occurs with probability $P_{\mbox{\scriptsize s}}=
p^2 |\alpha_1^{(L_1)}|^2/2 \approx 0.1620\times p^2$, 
cf.\ (\ref{Rcoeff}). The maximum
success probability, achievable by choosing suitable multiples of
length $L_0$, is given by  $P_{\mbox{\scriptsize
    s}}^{\mbox{\scriptsize max}}\approx 0.1667
\times p^2$.             

\pstex{h!}{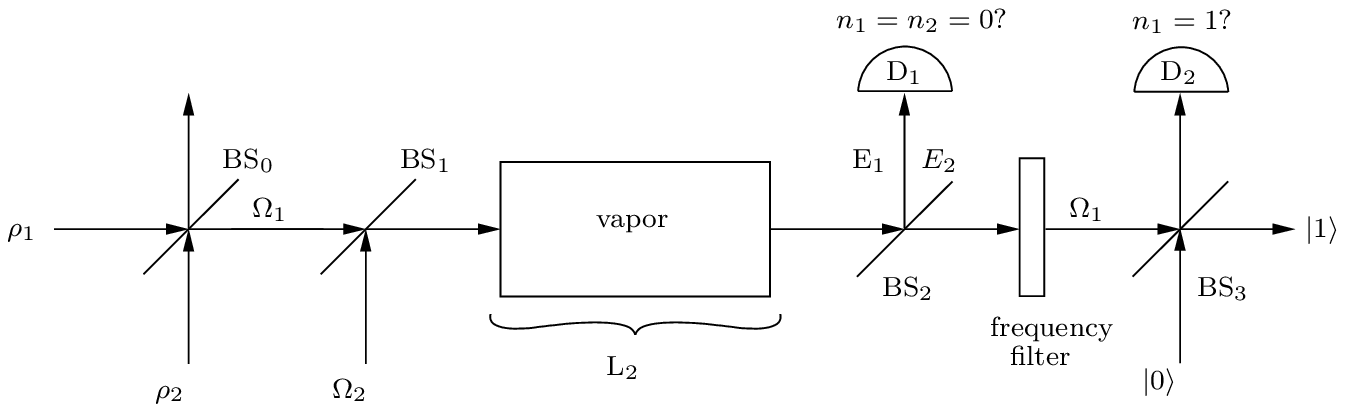}{Setup with
  nonlinear medium (vapor) of length $L_2=2L_0/3$. Detector $D_1$
is   conditioned on  zero photons in $E_1$ and $E_2$. The
  success probability is $P_{\mbox{\scriptsize s}}\approx 0.2291\times p^2$.}{FigB}

The second alternative to prepare the  single-photon state $\ket{1}$ using the
JF-setup is sketched in Fig.\ \ref{FigB}. This time the length of the
medium (vapor) is given by $L_2=3L_0/2$, which
induces the state change 
\begin{align}
&\ket{0,0,0}\xrightarrow{U(t=L_2/c)} \ket{0,0,0}\nonumber\;,\\
&\ket{1,0,0}\xrightarrow{U(t=L_2/c)} i\ket{0,1,1}\;,\nonumber\\
&\ket{2,0,0}\xrightarrow{U(t=L_2/c)} \alpha_0^{(L_2)}\ket{0,2,2}+\alpha_1^{(L_2)}\ket{1,1,1}
+ \beta^{(L_2)}\ket{2,0,0}\,.
\end{align}
After  detection of zero photons in detector D$_1$ (cf.\ Fig.\
\ref{FigB}), $\Omega_1$ either contains zero or two photons. Detecting
one photon in one of the outputs of a subsequent
symmetric beam splitter indicates one photon in the other output. 
These measurement outcomes and thus the single photon generation
occurs with probability $P_{\mbox{\scriptsize s}}=p^2|\beta^{(L_2)}|^2/4\approx
0.2291\times p^2$. By varying $M\in\mathbb{N}$ in $L_2=(2M-1)L_0/2$ one can achieve 
$P_{\mbox{\scriptsize s}}^{\mbox{\scriptsize max}}=0.25\times p^2$. 

The latter implementation generates a single photon with higher
probability, but it requires an additional detector.  The two alternatives
possess higher success probabilities than the implementation of our
original scheme  by means of the JF-mechanism ($P_{\mbox{\scriptsize
    s}}
\approx 0.0234$).  
However, the alternative devices depend  on the feasibility of the 
JF-mechanism, while our original scheme may be realized in many ways, i.e., by
employing different TPAMs. Moreover,  our original scheme as
described in Sections~\ref{Scheme} and \ref{Calculation} can, 
in principle, yield higher success probabilities. Please note 
that the success probabilities mentioned above can be doubled 
by processing the second output of BS$_0$ in the very same 
manner as the first output (cf.~to ~Sec.~\ref{Calculation}).

\end{document}